\documentclass[12pt]{article}

\usepackage[utf8]{inputenc}
\usepackage{geometry}
\usepackage{setspace}
\usepackage{apacite}
\usepackage{natbib}
\usepackage{graphicx}
\usepackage{lineno} 
\usepackage{authblk}

\usepackage{multirow}
\usepackage{amsmath,amssymb,amsfonts}
\usepackage{amsthm}
\usepackage{mathrsfs}
\usepackage{xcolor}
\usepackage{textcomp}
\usepackage{manyfoot}
\usepackage{booktabs}
\usepackage{algorithm}
\usepackage{algorithmicx}
\usepackage{algpseudocode}
\usepackage{listings}
\usepackage{pgffor}
\usepackage{tikz}
\usepackage{subcaption}
\usepackage{float}

\definecolor{darkgreen}{RGB}{0, 100, 0}
\definecolor{rpurple}{RGB}{160, 32, 240}

\newcommand{\blue}{\raisebox{-0.9pt}{\tikz{\fill[blue](0,0) rectangle (2mm,2mm);}}}
\newcommand{\red}{\raisebox{-0.9pt}{\tikz{\fill[red](0,0) rectangle (2mm,2mm);}}}
\newcommand{\green}{\raisebox{-0.9pt}{\tikz{\fill[darkgreen](0,0) rectangle (2mm,2mm);}}}
\newcommand{\purple}{\raisebox{-0.9pt}{\tikz{\fill[rpurple](0,0) rectangle (2mm,2mm);}}}

\newtheorem{theorem}{Theorem}

\title{Bringing Age Back In: Accounting for Population Age Distribution in Forecasting Migration}
\author[1]{Nathan G. Welch}
\author[2]{Hana \v{S}ev\v{c}\'{i}kov\'{a}}
\author[3]{Adrian E. Raftery}

\medskip
\affil[1]{\small Department of Statistics, \textbf{University of Washington},\par Box 354322, Seattle, WA 98195-4322, USA\par Email: nwelch@uw.edu}

\affil[2]{\small Center for Statistics and the Social Sciences, \textbf{University of Washington},\par Box 354322, Seattle, WA 98195-4320, USA\par Email: hanas@uw.edu}

\affil[3]{\small Departments of Statistics and Sociology, \textbf{University of Washington},\par Box 354322, Seattle, WA 98195-4322, USA\par Email: raftery@uw.edu}

\date{\today}

\begin{document}

\maketitle
\thispagestyle{empty} 

\setcounter{page}{1} 


\section*{Abstract}
The link between age and migration propensity is long established, 
but existing models of country-level net migration 
ignore the effect of population age distribution on past and projected 
migration rates. We propose a method to estimate and forecast 
international net migration rates for the 200 most populous countries, 
taking account of changes in population age structure. 
We use age-standardized estimates of country-level net migration rates and 
in-migration rates over quinquennial periods from 1990 through 2020 to 
decompose past net migration rates into in-migration rates and 
out-migration rates. 
We then recalculate historic migration rates on a scale that removes the 
influence of the population age distribution. This is done by scaling past 
and projected migration rates in terms of a reference population and period. 
We show that this can be done very simply, using a quantity we call the {\it migration age structure index} (MASI).
We use a Bayesian hierarchical model to generate joint probabilistic forecasts of 
total and age- and sex- specific net migration rates over five-year periods 
for all countries from 2020 through 2100. 
We find that accounting for population age structure in historic and forecast net migration 
rates leads to narrower prediction intervals by the end of the century for 
most countries. Also, applying a Rogers \& Castro-like migration age 
schedule to migration outflows reduces uncertainty in population pyramid forecasts.
Finally, accounting for population age structure leads to less out-migration among 
countries with rapidly aging populations that are forecast to contract most rapidly
by the end of the century. This leads to less drastic population declines than are 
forecast without accounting for population age structure.

\medskip
\begin{center}
\textbf{keywords: }\textit{migration, probabilistic, population, projection, forecast}
\end{center}


\section{Introduction}\label{Introduction}

Preparations for the next great demographic era are underway around the 
world \citep{kim2019,dni2021}. 
Decades of declining and then sub-replacement fertility 
created a demographic dividend, where the share of the working-age
population far outweighed the share of children and older adults no longer
participating in the workforce \citep{bongaarts2009}. 
Now the average ages of populations around the world are increasing
at the fastest rate in history \citep{imf2022}.  
International migration may blunt the 
impact of this dynamic, but migration alone will not fully mitigate the 
realities of the next demographic era \citep{rand2005,coleman2008}.
Accounting for past and forecast population age structure reveals that 
migration may be an even less effective mitigation than currently understood 
\citep{munz2013,lee2011}. 
However, existing forecasting methods fail to account for this eventuality, 
leaving the role of population age structure in migration dynamics to imprecise 
qualitative conjecture. 

In the period from 2000 through 2015, growth in the total number of migrants in 
developed countries averaged 3.0\% annually, outpacing the 0.6\% annual 
population growth in these nations \citep{mgi2016}. 
More robust forecasts of migration dynamics are critical to the long-term 
policymaking process as countries compete for the same migrants to mitigate
the impacts of sustained sub-replacement birth rates. 
In the coming decades, the portfolio of migrant supplying countries will have to evolve
as birth rates fall, origin populations age, and the share of migration-age people in 
historic migrant supplying countries falls. 

We propose a probabilistic net migration forecasting approach that accounts 
for the impacts that population age structure had on historic migration 
rate estimates.  We use these to generate new probabilistic population forecasts
for all countries. 
The resulting forecasts quantify how different the global population 
distribution could be after accounting for past and projected shifts in 
the population age structure. 

The article is organized as follows.
Section \ref{Background} describes the background to our study.
Section \ref{Data} summarizes the data and methods used to generate age 
standardized migration forecasts. 
It details each step of the population forecasting method. 
This is followed in Section \ref{Validation} by an evaluation of the age-standardized migration method 
compared to forecasts that do not account for the role of population age 
structure in population forecasting. 
Then Section \ref{Results} summarizes differences between population forecasts with and 
without accounting for population age structure in the migration component 
of the forecasts. 
We conclude in Section \ref{Discussion} with a discussion of the contributions 
and limitations of our approach.

\section{Background}\label{Background}

The association between migration and age is long established. 
\citet{rogers1981} identified a consistent pattern in the age profile of 
out-migrants and proposed a model migration age schedule.
Even though modern migration estimation and forecasting methods use 
migration age schedules to account for the impact of migration 
on the sending and receiving populations, population age structure is rarely
an explicit consideration in estimation or forecasting. 
This is reasonable over short-term forecast horizons given the relatively 
slow pace of population age structure changes, but less so over the long term.

\citet{kupiszewski2002} discussed differences in migration 
forecasts when the sending population age-structure is ignored. 
Specifically, it was argued that ignoring the limited number 
of migration-age individuals in Poland and ignoring natural population aging 
contributed to unrealistically large out-migration forecasts from Poland to 
other European Union (EU) countries before accession to the EU in 2004.  

\citet{fertig2005} argued that neglecting population age in calculating 
migration rates obscures the underlying population dynamics. 
They proposed an adjusted net migration rate that is calculated 
as the number of migrants aged 0-39 divided by the sending country population
aged 0-39.     
This approach reduces distortions created by the population of older people who 
tend to make up a relatively small proportion of migrants, but effectively sets 
older age migration to zero and ignores the age structure of the critical
0-39 age group.

\citet{kolk2019} reviewed age-specific migration and the analogy to
 age-specific and total fertility rate definitions.  
It was argued that population-level measures should be adopted for both
subnational and international migration estimation. 
The authors outlined key challenges to extending such methods to an 
international context, specifically properly specifying the population 
at risk of in-migration and differences in the definition of an 
international migrant from country to country. 

Fully probabilistic population forecasting is an active area of demographic research \citep[e.g.][]{hyndman2008,raftery2012,raftery2014,wisniowski2015,yu2023}.
\citet{azose2016} developed a probabilistic net migration model and used it to produce probabilistic population forecasts for all countries through 2100.
\citet{welch2022} proposed a fully probabilistic population forecasting method that 
uses a probabilistic model of bilateral migration flows for all countries through 2045. 
However, none of these population forecasting methods systematically 
account for population age structure changes in both the historic data used 
to fit these models and in forecasting. 

\citet{raftery2023} proposed a method for accounting for population age
structure in probabilistic migration and population forecasts, and 
applied it to very long term population forecasts, to 2300, motivated
by the problem of estimating the social cost of carbon.
Like us, they used net migration data, which has advantages of
data availability and analytical simplicity over methods based on more 
complete migration data, such as between-country flows or in- and out-migration
flows for countries. 
However, a problem with this is that it does not take account of the different age distributions of in- and out-migrants \citep{rogers1990}. 
They avoided this problem by assuming
that when a country experienced net in-migration in a period, it was all 
in-migration and there was no out-migration, and conversely for out-migration.
Here we develop a new method that avoids this unrealistic assumption while 
still using net migration data.

\section{Data \& Methods}\label{Data}

\subsection*{Data}

Recent methodological innovations make it possible to estimate globally consistent 
country-level total migration inflows and outflows from Census counts of the
number of people by country of birth in all countries 
\citep{abel2013,azose2019}. 
\citet{abel2019} found that the pseudo-Bayes method of \citet{azose2019} to estimate bilateral migration flows
led to the most accurate migration estimates among several extant methods. 
Unfortunately, pseudo-Bayes migrant flow estimates over five-year periods are available 
for only six five-year periods starting in 1990 and ending in 2020 \citep{abel2019}. 
Estimating and forecasting bilateral migration flows has a number of 
advantages \citep{welch2022}, but population age structure changes 
too slowly to solely rely on these estimates for long-term migration forecasting. 

The United Nation's World Population Prospects (WPP) publishes globally consistent 
net migration estimates back to the 1950-1955 period \citep{unwpp2022}. 
The net migration rate is defined as the difference between migration
inflows and outflows divided by population size. 
As such, the population age structure at the given time is not accounted for.

The pseudo-Bayes flow estimates and net migration rate estimates have independent value. 
The pseudo-Bayes flow estimates contain more detailed information in the form of bilateral flows, 
but the length of such time series is limited \citep{abel2019}. 
On the other hand, 
the availability of net migration rate estimates going back to 1950 provides 
a measure of migration over many more historic periods \citep{unwpp2022}. 
We use strengths of both data sets to develop a new combined data series. 
Migration flow estimates are calculated with respect to specific revisions of the WPP. 
The most recent flow estimates were estimated using WPP 2019 \citep{abel2019}. 
We use WPP 2019 \citep{unwpp2019} and the associated flow estimates for all analyses. 

We fit a model, described in Section~\ref{Estimation}, to the 1990-2020 period from which both in-migration and net migration 
rates can be estimated for all countries with populations of 100,000 or more.
This model is then used to approximate the contribution that in-migration made to the 
net migration rate prior to 1990. 
Our model-based approach is an efficient solution that allows for the estimation of 
globally consistent age-standardized in-migration and out-migration rates for five-year 
periods starting in 1950 and running through 2020.
The successful decomposition of net migration rates into in-migration and out-migration 
rates is critical to calculating migration rates adjusted for the population age structure 
that gave rise to those rates. 

\subsection*{Age Standardization of Net Migration Rates}\label{Estimation}
For reasons of data availability across all countries, and analytic simplicity,
the UN uses all-age net migration as part of the basis for its
population projections for all countries \citep{unwpp2019}.
They project this forwards deterministically, combining it with probabilistic
projections of fertility and mortality, to obtain probabilistic
population projections conditional on the projected levels of future migration.
As an alternative, \citet{azose2015} proposed a probabilistic approach to 
forecasting future net migration rates (i.e. net migration divided by population).
Neither the UN's deterministic approach nor the probabilistic approach
of \citet{azose2015} takes account of historic or future changes 
in population age structure. 

We propose to modify the approach of \citet{azose2015} to account for
population age structure. We do this by creating an age-standardized version
of the net migration rate. This is tricky, because historic net migration data
are not disaggregated by age. Also, net migration is equal to in-migration
minus out-migration, and the age profiles of in- and out-migration can 
differ. As pointed out by \citet{rogers1990}, this makes it hard to 
standardize net migration rates directly, and indeed this was one of the
arguments of \citet{rogers1990} for not using net migration at all. 
Here we propose a method for doing so which hopefully reduces these concerns.

We start by developing a method for decomposing historic net migration into
in- and out-migration. This uses a subset of the data for which estimates
of both in- and out-migration are available (i.e.~1990-2020), and uses this 
to estimate in- and out-migration for the entire data period (i.e.~1950-2020).
We then develop a method for age-standardizing out-migration. This turns
out to be very simple, relying on a quantity we call the {\it migration
age structure index} (MASI). 
We extend this to in-migration, and hence to net migration. 
Finally we apply the probabilistic forecasting method of 
\citet{azose2015} to the age-standardized net migration rates, and input
the resulting forecasts  to the overall probabilistic population projection 
method. The notation used in describing our method is shown in 
Table \ref{tab:notation}.

\begin{table}[htb]
    \caption{Notation for migration age-standardized migration method}
    {
    \small
    \begin{tabular}{ll}
    $O_{i,t}$ & Integer-valued outflow from origin $i$ in period starting with year $t$\\
    $O_{i,t,a}$ & Integer-valued outflow from origin $i$ in period starting with year $t$ \\
        & of age group $a$-$(a+4)$ \\
    $I_{i,t}$ & Integer-valued inflow to destination $i$ in period starting with year $t$\\
    $N_{i,t}$ & Integer-valued net migrant flow in country $i$ in period starting with year $t$\\
    $P_{i,t,a,s}$ & Integer-valued population of origin $i$ at the start of period $t$ for age group\\
        & $a$-$(a+4)$ and sex $s \in \{\text{male}, \text{female}\}$ \\
    $P_{i,t,+,+}$ & Integer-valued population of origin $i$ at the start of period $t$ \\
    $\tilde{P}_{i,t,+,+}$ & Integer-valued population of origin $i$ at risk of migration over period $t$ to $t+5$, \\
        & defined as $P_{i,t+5,+,+} - N_{i,t}$\\
    $\mbox{OMR}_{i,t}$ & $= O_{i,t}/\tilde{P}_{i,t,+,+}$. Out-migration rate for country $i$ in the period starting in year $t$ \\
    $\mbox{OMR}_{i,t,a}$ & $= O_{i,t,a}/\tilde{P}_{i,t,a,+}$. Out-migration rate for country $i$, period starting in \\
        & year $t$ and age group $a$\\
    $G_{i,t}$ &  $= \sum_a \mbox{OMR}_{i,t,a}$. Gross Migraproduction Rate (GMR) for country $i$ and period\\ 
        & starting in year $t$ \\
    $\mbox{IMR}_{i,t}$ & $= I_{i,t}/\tilde{P}_{i,t,+,+}$. In-migration rate for period starting in year $t$ \& country $i$\\
    $\mbox{NMR}_{i,t}$ & $= \mbox{IMR}_{i,t} - \mbox{OMR}_{i,t}$. Net migration rate for country $i$ \& period $t$\\
    $\pi_{i,t,a}$ & $= \tilde{P}_{i,t,a,+}/\sum_a \tilde{P}_{i,t,a,+}$. Share of population from age group $a$ in period $t$ \\
    $\pi^*_{i,t,a}$ & Reference population age distribution for country $i$ in period $t$ \\
    $R_a$ & Reference migration age schedule normalized so that $\sum_a R_a = 1$\\
    $C_{i,t}$ & $= \sum_a R_a \pi_{i,t,a}$. The migration age structure index (MASI), a scalar accounting \\
        & for the population and migrant age structure  in the migration rate at\\
        & time $t$ for country $i$\\
    $\check{C}_{t}$ & $= \sum_{i,a} R_a \pi_{i,t,a}$. MASI for the world at time $t$ \\
    $\mbox{OMR}_{i,t}^{\star}$ & $= \mbox{OMR}_{i,t} \times C_{i,\text{baseline}}/C_{i,t}$. Age-standardized out-migration rate\\
    $\mbox{IMR}_{i,t}^{\star}$ & $= \mbox{IMR}_{i,t} \times \check{C}_{\text{baseline}}/\check{C}_{t}$. Age-standardized in-migration rate\\
    $\mbox{NMR}_{i,t}^{\star}$ & $= \mbox{IMR}_{i,t}^{\star} - \mbox{OMR}_{i,t}^{\star}$. Age-standardized net migration rate
    \end{tabular}
    }
    \label{tab:notation}
\end{table}

\subsubsection*{Decomposing Historic Net Migration into In- and Out-migration}
Net migration age-standardization takes place on the 
inflow and outflow components of the net migration rate. 
Pseudo-Bayes estimates \citep{azose2019,abel2019} of migration flows are available over five-year periods starting in 1990 and 
ending in 2020 (6 periods). 
Net migration rate estimates are available for five-year periods back to 1950 \citep{unwpp2019}.  
We use a mixed-effects model to relate the 1990-2020 in-migration rate, 
$\mbox{IMR}_{i,t}$, to the net migration rate, $\mbox{NMR}_{i,t}$, 
for each country $i$ over these six five-year periods, $t$, from 1990 through 2020. 
Our mixed-effects model is defined as follows:
\begin{align} 
\mbox{IMR}_{i,t} &= \beta_{0,i} + \beta_1 \max \left( \mbox{NMR}_{i,t}, 0 \right) + \varepsilon_{i,t} ,  \nonumber \\
\beta_{0,i} &\sim \text{Normal}\left(\beta_0, \sigma^2_{between} \right) , \nonumber \\
\varepsilon_{i,t} &\sim \text{Normal}\left(0, \sigma^2_{within} \right).
\label{eq:mixedEffectsModel}
\end{align} 
This model uses a random intercept term to account for the average country-specific
in-migration rate associated with the net migration rate for each period from 1990-2020. 
Country intercepts, $\beta_{0,i}$, are concentrated around a global mean intercept, $\beta_0$. 
The difference in a country's random intercept from the global mean, $\beta_0$, 
is determined by the average difference from the global mean for that country in all periods. 
Outflow rates are implicitly defined by this model. 
Table \ref{tab:notation} defines each term in the model. 

The random intercept model was fit using the linear mixed-effects model implementation in the \textit{lme4} 
R package implementation \citep{lme4}.
Starting with $\mbox{NMR}_{i,t}$, the estimated in-migration rate is 
$\mbox{IMR}_{i,t} = \beta_{0,i} + \beta_1 \max \left( \mbox{NMR}_{i,t}, 0 \right)$. 
The net migration rate and estimated in-migration rate are then used to 
calculate the out-migration rate, $\mbox{OMR}_{i,t} = \mbox{IMR}_{i,t} - \mbox{NMR}_{i,t}$.

\begin{figure}[htb]
    \centering
    \includegraphics[width=\textwidth]{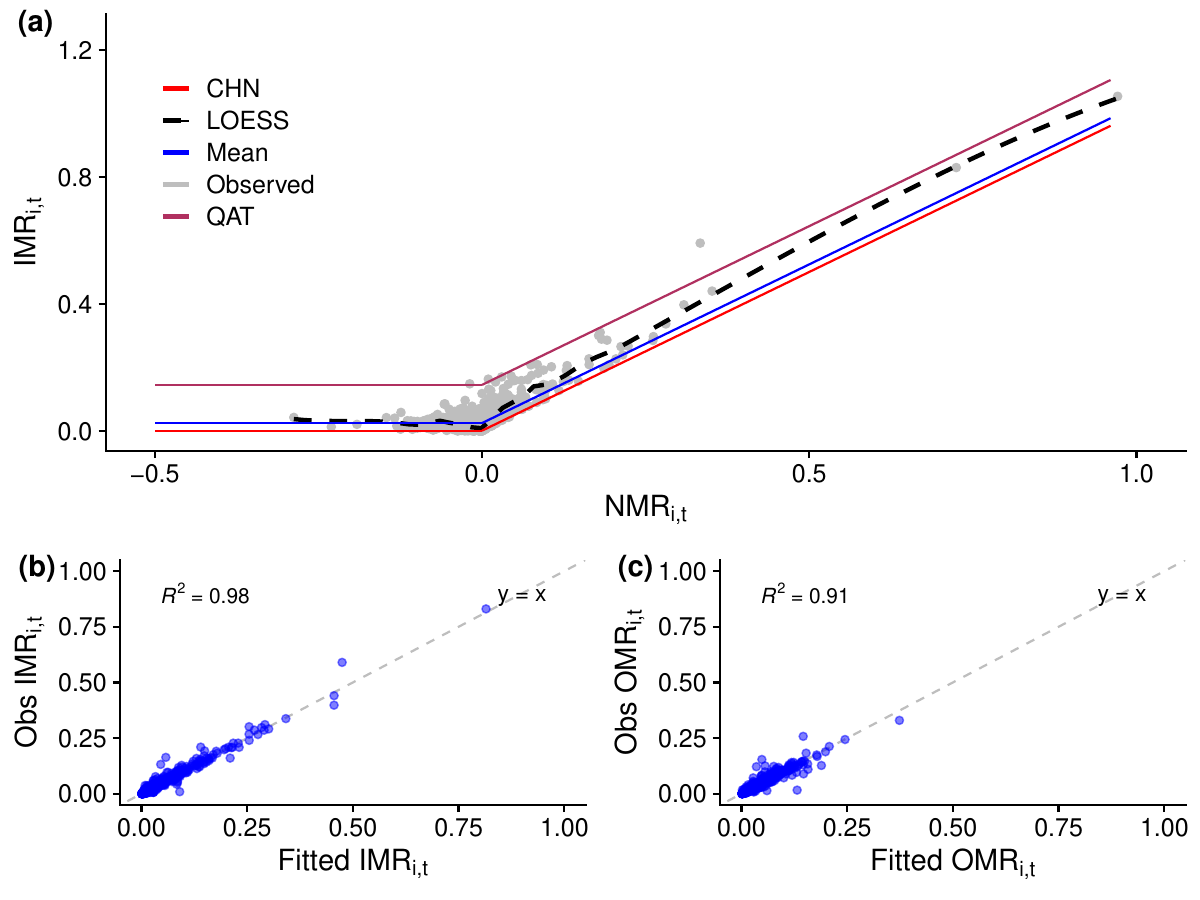} 
    \caption{(a) Observed in-migration rates versus net migration rates for all countries (points) 
    with mean model (blue), LOESS line (black dashed), and country-specific models for China 
    (CHN, red) and Qatar (QAT, maroon); (b) Observed in-migration rates versus fitted in-migration 
    rates compared to $y=x$ (dashed line); (c) Observed out-migration rates versus fitted out-migration 
    rates compared to $y=x$ (dashed line) for five-year-periods from 1990-2020.}
    \label{fig:rateScatter}
\end{figure}

Figure \ref{fig:rateScatter} summarizes the 1990-2020 data, model fit, and 
association between estimated and actual country-level migration flow rates.
Points in Figure \ref{fig:rateScatter}(a) show the observed $\max \left( \mbox{NMR}_{i,t}, 0 \right)$ 
on the horizontal axis and $\mbox{IMR}_{i,t}$ on the vertical axis. 
The discontinuity at $\mbox{NMR}_{i,t}=0$ reflects the fact that the in-migration rate 
must be non-negative. 
In the $\mbox{NMR}_{i,t} \le 0$ portion of the domain, the country-specific intercept, 
$\beta_{0,i}$, of model \ref{eq:mixedEffectsModel} establishes the minimum average magnitude of 
in-migration corresponding to a negative net migration rate for that country. 
When $\mbox{NMR}_{i,t} > 0$, the model in-migration rate increases from this minimum average 
inflow at a rate of $\beta_1$ per unit increase in $\mbox{NMR}_{i,t}$. 
The mean model in-migration rate is shown by the blue line in Figure \ref{fig:rateScatter}(a).

Many observations in Figure \ref{fig:rateScatter}(a) are highly concentrated around some values. 
The LOESS curve helps clarify the location of points that fall too close to each other to be seen separately. 
In the $\mbox{NMR}_{i,t} < 0$ portion of the domain, the similarity between the LOESS line 
and the $\beta_{0,i}$ values shows that the mean intercept term provides an accurate summary of the observed in-migration and net migration rates. 
Near $\mbox{NMR}_{i,t} = 0$, however, the disagreement between the model mean and LOESS 
line shows that many $\mbox{IMR}_{i,t}$ observations near $\mbox{NMR}_{i,t}=0$ could be 
overstated with the model mean alone. 
The LOESS line departs from the mean model most for large positive values of $\mbox{NMR}_{i,t}$. 
This is primarily due to the influence of a few large positive $\mbox{NMR}_{i,t}$ observations for 
a small number of countries. 

Finally, Figure \ref{fig:rateScatter}(a) shows two examples of country-specific average association 
between $\mbox{IMR}_{i,t}$ and $\mbox{NMR}_{i,t}$. 
The top line shows the model fit for Qatar (maroon), which experienced some of the highest 
migration rates in the world since 1990. 
The bottom line shows the model fit for China, a country with stable net out-migration 
over the same period. 
The differences between these two examples highlight the need for a model flexible enough to 
account for such wide variation between countries. 

Figure \ref{fig:rateScatter}(b) shows the fitted and observed $\mbox{IMR}_{i,t}$ for all 
countries from 1990-2020. 
The high level of agreement between fitted and observed in-migration rates with an $R^2=0.98$
is notable considering the simplicity of model (\ref{eq:mixedEffectsModel}). 
The agreement between fitted and observed out-migration rates in Figure \ref{fig:rateScatter}(c)
is lower with $R^2=0.91$, but is still strong. 
Taken together, Figures \ref{fig:rateScatter}(b)-(c) 
show that our model-based decomposition of the net migration rate is reasonable for the 
periods where net migration, in-migration, and out-migration rates are all available. 

\begin{figure}[htb]
    \centering
    \includegraphics[width=0.9\textwidth]{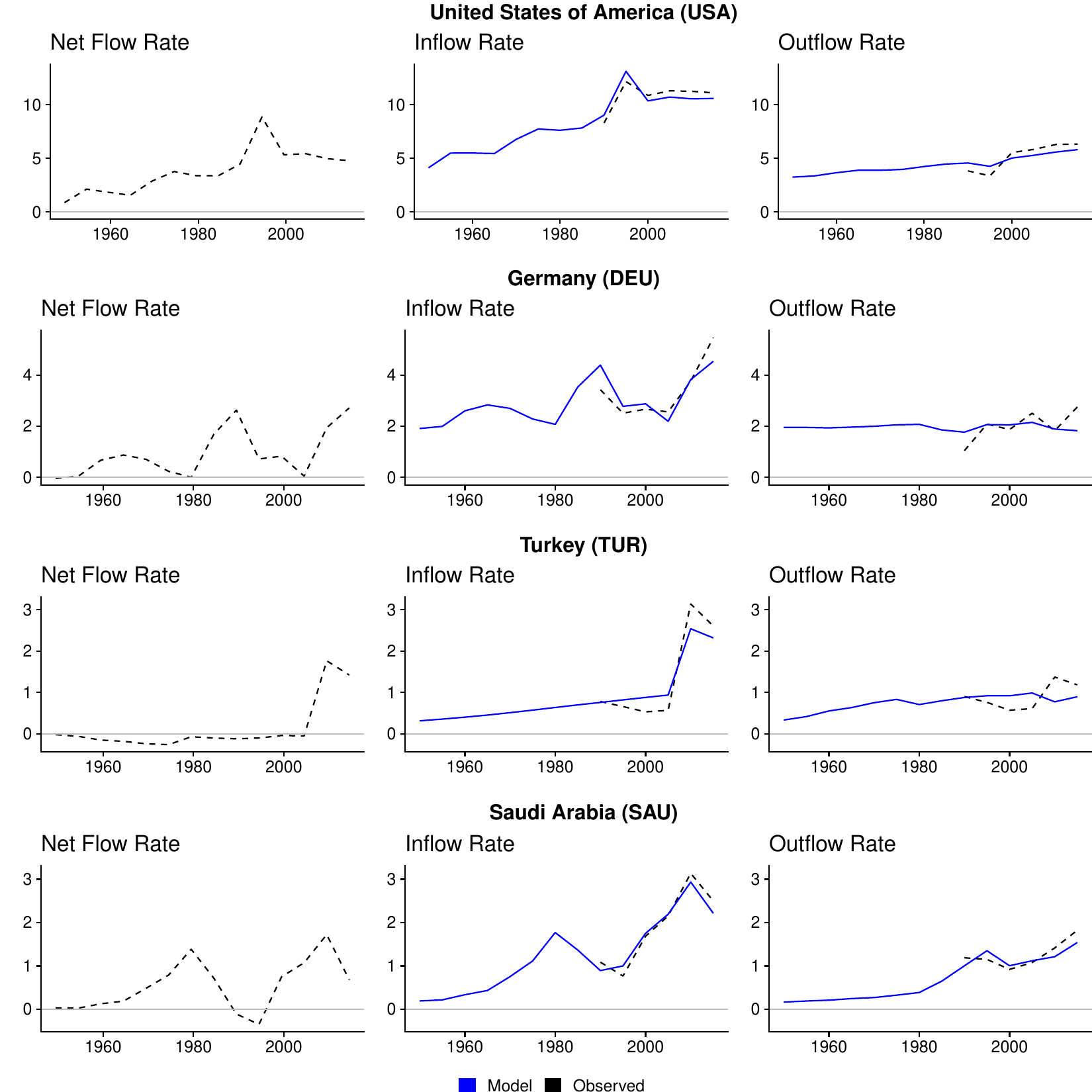} 
    \caption{Observed net migration (left column), decomposed into in-migration (middle column), and out-migration rates (right column), on the scale of annual migrants per thousand people,
    compared to mixed-effects model estimates for the United States, Germany, Turkey, and Saudi Arabia.
    Solid blue lines show the model-based estimates.
    Dashed black lines show the observed migration rates used for the estimation.}
    \label{fig:rateDecomposition}
\end{figure}

Figure \ref{fig:rateDecomposition} shows the observed net migration  
rate for five-year periods starting in 1950 and running through 2020 on the 
original scales. The inflow and outflow rates from 1990 through 2020 used for the estimation of our model
are shown in the second and third columns along with the mixed-effects model 
decomposition. 
Our model-based estimates of inflow and outflow rates applied to years prior to 1990 are also shown 
in the plots. 
Model-based estimates relating net migration rates to in-migration and 
out-migration rates are similar to the observed rates for 1990-2020. 
Estimated inflow and outflow rates prior to the 1990-1995 period appear plausible
for the periods where pseudo-Bayes estimates are available.

While the model-based decomposition is not perfect, Figure 
\ref{fig:rateDecomposition} shows that the mixed-effects net rate
decomposition approach leads to plausible estimates of in-migration and 
out-migration rates. 
The correlation between the observed and predicted rates was over 0.97. 
Broad agreement between observed inflow and outflow rates for 1990-2020 periods 
suggests that net migration rates prior to 1990 can be similarly decomposed into inflow 
and outflow rates. 

\citet{raymer2023} proposed an alternative approach to decomposing net migration into 
in-migration and out-migration.
This method uses a fixed constant to approximate the magnitude of net migration
attributed to total inflows and outflows in proportion to a country's population.
This method could also be used to decompose the net flow into total inflow and outflow 
in place of model (\ref{eq:mixedEffectsModel}) for a subset of countries or all countries
if desired. 
However, we found the proportion of the in-migration to net migration rates
using model (\ref{eq:mixedEffectsModel}) too variable to justify using a single constant
for all countries. 
The model (\ref{eq:mixedEffectsModel}) offers a systematic approach to specifying a 
quantity similar to the fixed constant used in \citet{raymer2023}, but one
that is country-specific, which our analyses suggest is needed.

\subsubsection*{Age Standardization of Out- and In-migration Rates}
The net migration rate, NMR$_{i,t}$, for country $i$ over 
period $t$ to $t+5$ is defined as the difference between the in-migration rate, 
$\mbox{IMR}_{i,t}$, and out-migration rate, $\mbox{OMR}_{i,t}$:
\begin{equation}
\mbox{NMR}_{i,t} = \mbox{IMR}_{i,t} - \mbox{OMR}_{i,t} = \frac{N_{i,t}}{\tilde{P}_{i,t,+,+}} = \frac{I_{i,t} - O_{i,t}}{\tilde{P}_{i,t,+,+}}.
\label{eq:netmigration}
\end{equation}
Here, $I_{i,t}$ denotes the total inflow count, $O_{i,t}$ the total outflow
count, and $N_{i,t}=I_{i,t}-O_{i,t}$ the total net flow count over the period 
starting in year $t$.
We define the denominator of the net migration rate, $\tilde{P}_{i,t,+,+} = P_{i,t+5,+,+}-N_{i,t}$, 
as the population of country $i$ at the end of the period starting in year 
$t$ before factoring in the change due to net migration. 
This denominator specifies the population at risk of migration over the period. 

Historic net migration rate estimates are not disaggregated by age, but we aim to 
standardize migration rates to remove the effects of population age structure
differences among countries in the same period and within countries across 
periods.
The age pattern of age-specific migration rates is known to be relatively consistent 
over time and country of origin \citep{rogers1981}, but the rate defined in 
equation (\ref{eq:netmigration}) obscures the influence of the sending and 
receiving country population age structures, as discussed by \citet{rogers1990}. 
A population age-standardized net migration rate should account for the 
age structure of the sending populations in both components of the net rate 
as the migrant age distribution and totals are primarily linked to the 
origin population age structure.

Let $\pi_{i,t,a}$ denote the population in age group $a$ as a proportion of the total population in country $i$ at time $t$, namely
\begin{equation}
\pi_{i,t,a} = \frac{\tilde{P}_{i,t,a}}{\sum_a \tilde{P}_{i,t,a}} \,.
\end{equation}
Since the origin population of inflows to country $i$ consists of every country other
than $i$ in period starting in year $t$, we approximate the proportion of the global population in age-group $a$ by
\begin{equation}
    \check{\pi}_{t,a} \approx \frac{\tilde{P}_{+,t,a,+}}{\tilde{P}_{+,t,+,+}} 
    \quad \text{where } \quad \tilde{P}_{+,t,a,+} = \sum_{i,s} \tilde{P}_{i,t,a,s}
    \quad \text{and} \quad \tilde{P}_{+,t,+,+} = \sum_{i,s,a} \tilde{P}_{i,t,a,s}\, .
\end{equation}

The out-migration rate for country $i$, period $t$ and age group $a$ is
\mbox{OMR}$_{i,t,a}$, and $G_{i,t} = \sum_a \mbox{OMR}_{i,t,a}$ is the
{\it Gross Migraproduction Rate} (GMR) for country $i$ and period $t$,
as defined by \citet{rogers1981}. This is a measure of overall migration
that is not affected by population age distribution, rather as the 
Total Fertility Rate (TFR) is a measure of overall fertility that is
independent of the age distribution of women. 

The values of OMR$_{i,t,a} / G_{i,t}$, which give the age pattern of
age-specific out-migration rates, tend to be stable over time and place,
reflecting a tendency for international migration to be largely concentrated
among people aged 15--35 and their dependent children, peaking in the twenties,
as pointed out by \citet{rogers1981}. They also proposed a parametric
model for this pattern, the famous Rogers-Castro curve. 
As an approximation, we thus consider the situation where this ratio is 
constant over time and space, so that OMR$_{i,t,a} / G_{i,t} = R_a$
for all $i,t$, where $\sum_a R_a = 1$. Under this assumption, the quantity $R_a$ is the same
for all countries and time periods, and could be modeled by a Rogers-Castro
curve, or estimated empirically.

We then have an exact result for the out-migration rate for a reference
population with a given age distribution: 

\begin{theorem}
Consider a population that has the same 
age-specific out-migration rates as country $i$ in period $t$, but
 a different population age distribution given by $\pi^*_{i,t,a}$. Then
this population has out-migration rate
\begin{equation}
\mbox{OMR}^*_{i,t} = \mbox{OMR}_{i,t}  
\frac{\sum_a \pi^*_{i,t,a} \mbox{OMR}_{i,t,a}}
 {\sum_a \pi_{i,t,a} \mbox{OMR}_{i,t,a}} .
\label{eq-OMR*}
\end{equation}
If both populations have the same age-specific pattern of 
migration rates, $R_a$, then
\begin{equation}
\mbox{OMR}^*_{i,t} = \mbox{OMR}_{i,t} \frac{C^*_{i,t}}{C_{i,t}},
\label{eq-OMR*.Ra}
\end{equation}
where $C_{i,t} = \sum_a \pi_{i,t,a} R_a$ is the migration age structure
index (MASI). 
\end{theorem}

\noindent The proof of this theorem can be found in the Appendix.

This indicates that one can age-standardize the out-migration rate to 
a given reference population using equation (\ref{eq-OMR*.Ra}). This is a
remarkably simple method, since it involves only a simple multiplication by
the MASI, $C_{i,t}$. The MASI involves only the population age
distribution and the invariant migration age-pattern $R_a$, but not the
age-specific migration rates themselves, which cancel.

We can age-standardize in-migration by viewing it as out-migration to 
country $i$ from the rest of the world. We approximate the age distribution
of the rest of the world by the age distribution of the world as a whole.
This yields an age-standardized in-migration rate, IMR$^*_{i,t}$. 
Finally, we obtain the age-standardized net migration rate as 
NMR$^*_{i,t} = \mbox{IMR}^*_{i,t} - \mbox{OMR}^*_{i,t}$. 

Typically the reference population age distribution will be the distribution
in a particular year. In this article, we standardize to the age pattern 
of 2020. 
To calculate the age-standardized in-migration and out-migration rate of 
period $t$ in terms of the 2020 population age structure, remove 
the population age structure effects in period $t$ and scale the migration 
rate in terms of the 2020 reference population:
\begin{align}
    \mbox{IMR}_{i,t}^{\star} &= \mbox{IMR}_{i,t} \frac{\check{C}_{2020}}{\check{C}_t} \nonumber \\
    \mbox{OMR}_{i,t}^{\star} &= \mbox{OMR}_{i,t} \frac{C_{i,2020}}{C_{i,t}}
\end{align} 
where $\check{C}_t$ denotes the MASI for the world at time $t$.

This implies that the age-standardized net migration rate is given by 
$\mbox{NMR}_{i,t}^{\star} = \mbox{IMR}_{i,t}^{\star} - \mbox{OMR}_{i,t}^{\star}$. 
This measure of migration puts historic and forecast rates of migration 
into the context of the 2020 population age structure, removing variation 
in migration rates attributable to population age structure differences 
from period to period. 
This specification also shows how to convert between the age-standardized rates
and the historic or future rates in terms of the reference population. 

After estimating historic inflow and outflow rates calculated from model 
(\ref{eq:mixedEffectsModel}), age-standardized inflow, outflow, 
and net flow rates can be calculated for forecasting. 

\subsubsection*{Net Migration Rate Model}
We use the resulting rates to develop a probabilistic model for age-adjusted net migration rates.
Using the above estimates of inflows and outflows, we compute the age-standardized net migration rates, 
$\mbox{NMR}_{i,t}^{\star}$, for all countries $i$ and for $t$ from 1950 through 2020, 
using the 2020 population age structures as the baseline population, as shown in Section~\ref{Estimation}.
We then fit the Bayesian hierarchical model of \cite{azose2015}.
Hyperparameter values for this model required no adjustment since the 
default specifications were broadly defined and our specification of the 
net rate is on the same scale as that of \cite{azose2015}. 
This is done using a Markov chain Monte Carlo (MCMC) algorithm.
This yields a sample of model parameters.

\subsection*{Forecasting}\label{Forecasting}

The goal here is to forecast migration and population probabilistically by generating jointly a set of future migration and population trajectories.
We use the same approach to forecasting fertility and mortality as the UN projections 
for the UN's 2019 {\it World Population Prospects} (WPP) \citep{unwpp2019}.
Our approach to forecasting migration builds on the methods of
\citet{azose2015} and \citet{azose2016}, but modifies them to adjust
for differences in population age distribution over time and between 
countries.

For each future trajectory $j \in \{1, \dots, J\}$, we independently project the population for each country $i$ by five-year 
age group $a$ through $a+5$ and sex $s$ for year $t+5$ without migration.
The resulting projection, $\tilde{P}_{i,t+5,a,s}^{(j)}$,  denotes a realization of the total population that would be observed 
in year $t+5$ had no one in the population migrated in or out of each country. 
This approximates the population at risk of migrating for the period $t$ to $t+5$, by age and sex. 
We then independently generate an age-standardized net migration rate for the period 
$t$ to $t+5$, $\mbox{NMR}_{i,t}^{\star (j)}$, using the parameter sample from the Bayesian hierarchical model of \citet{azose2015} estimated in the previous subsection. 
The sampled age-standardized net migration rate is then decomposed into 
the age-standardized in-migration rate, $\mbox{IMR}_{i,t}^{\star (j)}$, and age-standardized 
out-migration rate, $\mbox{OMR}_{i,t}^{\star (j)}$, using the mixed-effects model (\ref{eq:mixedEffectsModel}) that has been fit to 
age-standardized historical rates (i.e., $\mbox{IMR}^{\star}$ and $\mbox{OMR}^{\star}$). 
The resulting age-standardized coefficients are denoted by $\beta_{0}^{\star}$ and $\beta_1^{\star}$. 

The mean age-standardized in-migration rate for trajectory $j$ and country $i$ is 
calculated as 
\begin{equation}        
    \mbox{IMR}^{\star (j)}_{i,t} = \beta_{0,i}^{\star} + \beta_1^{\star} \max \left( \mbox{NMR}^{\star (j)}_{i,t}, 0 \right). 
\end{equation}
The corresponding out-migration rate for the period starting in year $t$ is then given by 
\begin{equation}        
    \mbox{OMR}^{\star (j)}_{i,t} = \mbox{IMR}^{\star (j)}_{i,t} - \mbox{NMR}^{\star (j)}_{i,t}. 
\end{equation}
Age-standardized inflow and outflow rates then need to be converted back to period-specific rates to 
calculate inflow and outflow counts corresponding to the projected population age structure:
\begin{align}
    \mbox{IMR}_{i,t}^{(j)} &= \mbox{IMR}_{i,t}^{\star (j)} \times \check{C}_{t}^{(j)}/\check{C}_{2020} , \nonumber \\
    \mbox{OMR}_{i,t}^{(j)} &= \mbox{OMR}_{i,t}^{\star (j)} \times C_{i,t}^{(j)}/C_{i,2020}.
\end{align}
In-migration rates then are converted to inflow counts by age and sex, $I_{i,t,a,s}^{(j)}$,
by multiplying the inflow rate by the total country population for trajectory $j$ at time $t$ 
and then applying a Rogers \& Castro-like migration age schedule. 
These counts are further disaggregated by sex in proportion to the males and females in the population. 
Out-migration counts by age and sex, $O_{i,t,a,s}^{(j)}$, are similarly calculated.

While a Rogers \& Castro-like age schedule is appropriate for out-migration in most countries, the 
migration schedules in Gulf Cooperation Council (GCC) member states (Bahrain, Kuwait, Oman, Qatar, Saudi Arabia, 
United Arab Emirates) call for a different approach.
In recent decades, GCC country migration has been dominated by large inflows of primarily male migrant workers.
This workforce supplies the labor required to power some of the globe's largest construction and infrastructure 
projects. 
These temporary workers typically enter GCC countries on visas valid for two years or less. 
The path from temporary worker to long-term resident or citizen is nearly nonexistent in GCC countries. 
As a result, the population age structure of GCC countries should be influenced by changes in the volume of 
migrant workers, but the age distribution of the population should remain relatively constant as worker visas 
expire, those workers leave the country, and new workers arrive to replace them or add to their ranks. 
To maintain the population age structure of the population due to temporary workers, we take a different approach to the migration age schedule.

Instead of applying the standardized model migration age schedule in GCC countries, we used a 
different schedule to maintain the foreign-worker-dominated age structure.  
This modification ensures that the population age distribution in GCC countries remain consistent 
with the distribution of migrant worker stocks replaced by departing foreign workers. 
The out-migration schedule was modified so that outflows are dominated by older working age migrants,
instead of the Rogers \& Castro-like schedule used for all other countries. 
After accounting for the age distribution of the population due to in-migration, the 
outflow migration age schedule was determined by the difference between the population 
age distribution and the normalized Roger \& Castro-like migration age schedule. 
This formulation effectively re-weights the outflow schedule so that outflows are composed of older workers 
aging out of prime working age groups (i.e., 15--65). 
Outflow rates in GCC countries have been much smaller than inflow rates over recent decades, but the population 
age distribution has changed little over this time. 
Hence, these inflow and outflow migration schedule modifications ensure that the migrant worker population 
remains in the prime working age range rather than having large groups of migrant workers
aging in place with the resident population.  
We found that these adjustments led to much more plausible forecasts of GCC migration and population 
dynamics through 2100 compared to the approach used for the rest of the world. 

It was necessary to rebalance global inflows and outflows each period 
to maintain global net-zero migration. 
\citet{azose2015} and \citet{azose2016} proposed methods for doing so,
but their methods  are not directly 
applicable to a forecasting method that produces inflow and outflow counts for each country. 
Adjusting net migration counts without specifying the proportion allocated to inflows or outflows 
leads to inconsistencies in the projected inflow and outflow totals compared to the net flow totals. 
We adjusted inflow and outflow totals by the global net overflow/underflow 
by adjusting inflow and outflow totals at the country-level in proportion to the country population as a share of the globe. 

The global rebalancing procedure is as follows. 
Let $\tilde{I}_{i,t,a,s}^{(j)}$ denote the $j^{\text{th}}$ inflow count trajectory for 
country $i$ for period starting in year $t$ for age group $a$ to $a+5$ and sex $s$.
Similarly, let $\tilde{N}_{i,t,a,s}^{(j)}$ denote the corresponding net migration count before global rebalancing. 
Then the adjusted inflow and outflow counts were calculated as
\begin{align}
    I_{i,t,a,s}^{(j)} &= \tilde{I}_{i,t,a,s}^{(j)} - w \left(\sum_{k \in K} \tilde{N}_{k,t,a,s}^{(j)} \right) \frac{\tilde{P}_{i,t,a,s}^{(j)} }{\sum_{k \in K} \tilde{P}_{k,t,a,s}^{(j)} } ,  \nonumber \\
    O_{i,t,a,s}^{(j)} &= \tilde{O}_{i,t,a,s}^{(j)} + (1-w) \left(\sum_{k \in K} \tilde{N}_{k,t,a,s}^{(j)} \right) \frac{\tilde{P}_{i,t,a,s}^{(j)} }{\sum_{k \in K} \tilde{P}_{k,t,a,s}^{(j)} } ,
\end{align}
with $w=0.5$, splitting the global net overflow/under-flow equally among inflows and outflows. 
The index $K$ defines two groups of countries used for normalization. 
Rebalancing was done among GCC countries (Bahrain, Kuwait, Oman, Qatar, Saudi Arabia, United Arab Emirates) 
and the labor supplying origin countries (Bangladesh, India, Indonesia, Philippines, Pakistan) in one group. 
The second group was composed of all other countries outside the GCC labor corridor. 
After this adjustment, the global net migration for each age group and sex equals zero and the net flow
of migrants for trajectory $j$ in country $i$ at time $t$ by age and sex becomes 
$N_{i,t,a,s}^{(j)} = I_{i,t,a,s}^{(j)} - O_{i,t,a,s}^{(j)}$.
As in \citet{azose2015}, these adjustments led to relatively small changes to forecasts. 

We then recalculated the balanced age-standardized net migration rate, $\mbox{NMR}_{i,t}^{\star(j)}$, as
\begin{equation}
    \mbox{NMR}_{i,t}^{\star(j)} = \left( \frac{I_{i,t,+,+}^{(j)}}{ \tilde{P}_{i,t+5,+,+}^{(j)} } \right) \times \check{C}_{2020}/\check{C}_{t}^{(j)} - \left( \frac{O_{i,t,+,+}^{(j)}}{ \tilde{P}_{i,t+5,+,+}^{(j)} } \right) \times C_{i,2020}/C_{i,t}^{(j)},
\end{equation}
which will be the rate jump-off for sampling in the next time period.
Finally, we account for net migration in the population projection generated at the start of the forecasting routine,
\begin{equation}
    P_{i,t+5,a,s}^{(j)} = \tilde{P}_{i,t+5,a,s}^{(j)} + N_{i,t,a,s}^{(j)} ,
\end{equation}
before moving to the next forecast period.

\section{Validation}\label{Validation}

\begin{figure}[htb]
    \centering
    \includegraphics[width=0.95\textwidth]{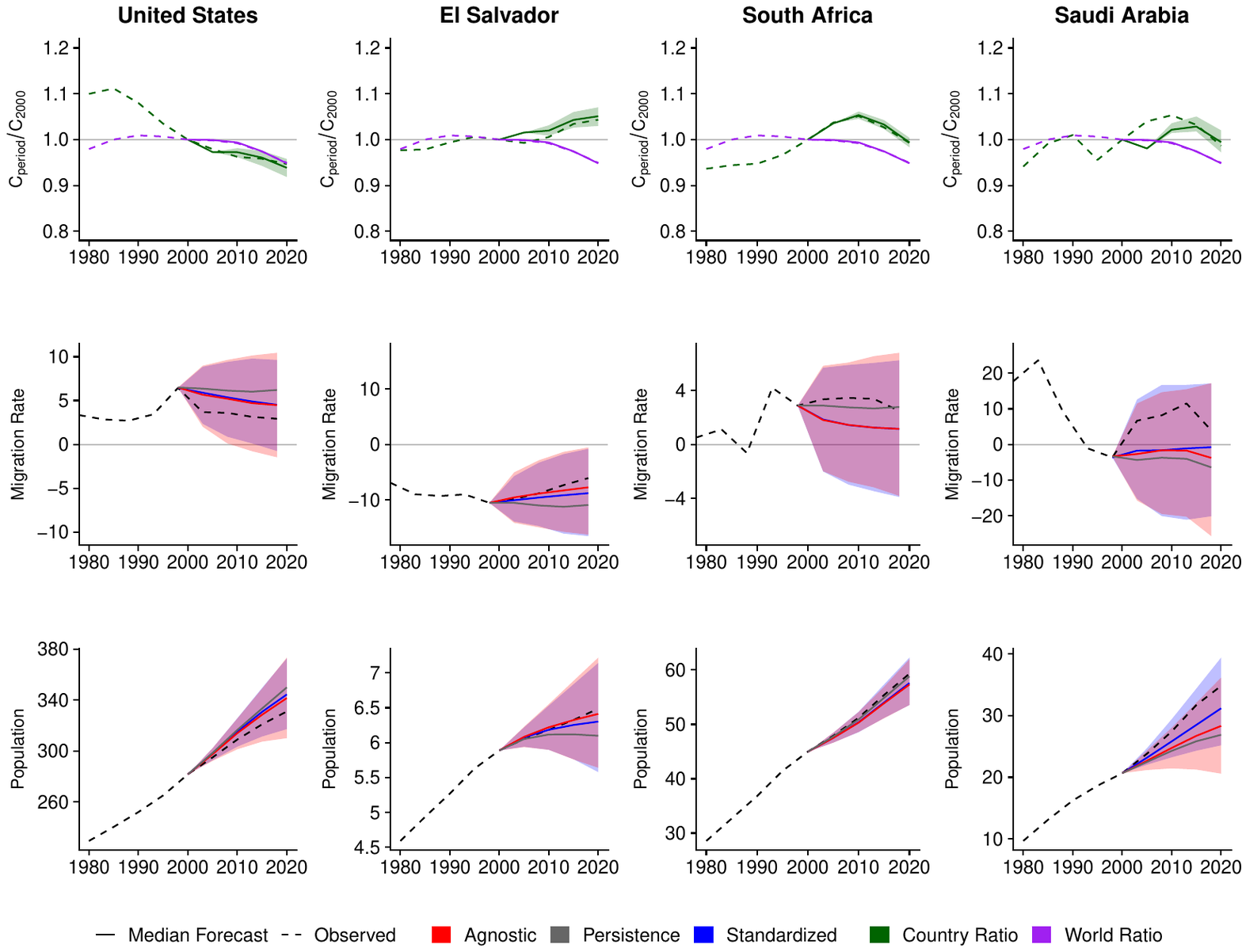} 
    \caption{ Migration age structure index (MASI) ratio for each country (\protect\green) and the globe (\protect\purple) 
    with base-year 2000, out-of-sample validation forecast of population (millions of people), 
    and age-standardized and age-agnostic net migration rate (net annual migrants per thousand), for four countries.
    Forecasts use probabilistic age-standardized net migration (\protect\blue), probabilistic age-agnostic net migration (\protect\red), 
    observed fertility, and observed mortality. Dashed lines in each plot indicate the observed values. 
    Solid lines indicate the median forecast. Shaded regions show the 80\% prediction interval. 
    Migration models were fit to 1950-2000 data. Forecasts are for the 2000-2005, 2005-2010, 2010-2015 and 2015-2020 periods. }
    \label{fig:validation}
\end{figure}

Age-standardized forecast efficacy was evaluated in terms of out-of-sample performance. 
Figure \ref{fig:validation} shows the observed and four-period migration and population
forecasts for four countries starting in the 2000-2005 period through the 2015-2020 period.
Columns show the United States, El Salvador, South Africa, and Saudi Arabia forecasts using 
1995-2000 migration rate persistence (effectively the current UN method) for every period, the age-agnostic method of \citet{azose2016}, 
and the age-standardized method described above.
Rows show the migration age structure index (MASI) ratio, i.e. the ratio of
the MASI for the period to the MASI for the baseline year 2000,  the net migration rate, and population for each country. 
Prediction intervals for both the age-agnostic and age-standardized method included the 
observed population and net migration rates in all cases. 
Population and migration forecasts using the age-standardized approach were as good
or better than the age-agnostic forecasts by the end of the period for all 
cases except in El Salvador in terms of agreement with the observed quantities. 
Persistence of 1995-2000 migration rates performed especially poorly for El Salvador and Saudi Arabia.

Age-standardized and age-agnostic forecasts largely agree with one another since population age
structure differences make little impact over such a short period of time.
Note that the proportion of the migration age population increased relative to the baseline population in El Salvador. 
The 0--15 year olds were the largest proportion of the population in 2000; however, the proportion of the population 
aged 0--15 fell during the validation period and the large 0--15 population cohort aged into the prime migration age cohorts by 
the end of the 2015-2020 period. 
As a result, El Salvador's age structure ratio increased from the 2000 baseline period.
The age structure ratio forecast outpaced the observed age structure ratio for the first two periods (2000-2010), but 
the prediction interval contained the observed age structure ratio for the last two periods (2010-2020).  

Global age ratio forecasts were well aligned with the true global age structure.
Country-level population age ratio forecasts showed mixed results, however.
Prediction intervals for the MASI ratio captured all or most of the observed values 
for the United States and South Africa. 
The forecast MASI ratio for Saudi Arabia was lower than 
the true value for the first two periods, but included the observed value by the end of the 
forecast period.
The forecast MASI ratio for El Salvador was consistently 
higher than the observed value from 2000 to 2010. 
These departures from the forecast are explained by the higher net migration observations 
compared to the forecasts for those countries. 

Global and country-level MASI ratio values equal one for the baseline population structure
(2000 in this case) by definition. 
When the global MASI is higher than the country-level MASI, then we expect
higher in-migration using the age-standardized method compared to the age-agnostic method. 
Conversely, when the country-level MASI ratio is below 1, we expect lower 
net out-migration compared to the baseline period using the age-standardized method and 
a model that does not account for population age structure. 
The MASI for the United States was below the global MASI and the 
baseline MASI value for the whole forecast period. 
The United States' MASI ratio also indicated an aging population 
compared to the baseline population age structure. 
Both of these factors should contribute to less out-migration from the United States as 
the global population had a net positive supply of migration-age people and 
the United States had a shrinking net positive supply of migration age people compared to the 
baseline population. 
Indeed, the age-standardized net migration forecast for the United States was shifted towards 
higher net in-migration compared to the age-agnostic forecast. 

The forecast MASI ratios for El Salvador and for South Africa were higher than 
the global MASI ratios  and above the baseline period, indicating that the age-standardized 
migration forecast should be shifted towards more net out-migration compared to the 
age-agnostic forecast. 
The age-standardized net migration rate was indeed more negative than the age-agnostic 
migration rate for El Salvador. 
However, age-standardized and age-agnostic net migration forecasts were indistinguishable 
for South Africa given the relatively small share of population change attributable to migration. 

The forecast MASI ratio for Saudi Arabia was higher than the global MASI ratio and 
above the baseline period, indicating that the age-standardized migration forecast 
should be shifted towards more net out-migration compared to the age-agnostic forecast
if the forecasting methods were otherwise identical. 
However, the age-standardized approach to forecasting Gulf Cooperation Council (GCC) 
uses a different approach to GCC net migration forecasting that makes the age-standardized 
and age-agnostic comparisons less relevant in this case. 
Still, net migration forecasts generated from the age-standardized method were more similar
to the observed rates. 
Since migration is the main contributor to population change in GCC countries, the small 
improvement in the age-standardized forecast led to a substantial improvement in the population
forecast compared to the age-agnostic approach. 

Multiple other forecast horizons were also evaluated. 
Predictions were generated one to four periods ahead of the last observed data.
Out-of-sample forecasts use the last observed population age structure as the baseline, 
fit each model to data available prior to the first forecast period, and generate forecasts 
for each period through the 2015-2020 period. 
We generated one-period-ahead forecasts for periods starting in 2000, 2005, 2010, and 2015. 
Two-period-ahead forecasts were generated for periods starting in 2000, 2005, and 2010. 
Three-period-ahead forecasts were generated for periods starting in 2000 and 2005. 
The four-period-ahead forecast was generated for the period starting in 2000 only. 
We did not evaluate the five-period-ahead forecast as it is not possible to fit the 
mixed-effects model (\ref{eq:mixedEffectsModel}) with only one period of inflow and 
outflow data. 

We evaluated the point forecasts in terms of the Mean Absolute Error (MAE),
the Log Mean Absolute Error (LMAE), and the Mean Absolute Scaled Error (MASE).
The LMAE is defined as follows:
\begin{align}
    l(y) &= \text{sign}(y) \left[ \log\left( \left| y \right| + c \right) - \log c \right] \text{ with } c>0  \nonumber \\
    \text{sign}(y) &= \begin{cases} 1 & y > 0 \\ 0 & y=0 \\ -1 & y<0 \end{cases}   \nonumber \\
    LMAE_t & = \frac{1}{200} \sum_{i=1}^{200} \left| l(f_{i,t}) - l(r_{i,t}) \right|. 
\end{align}
We used $c=1$.
The LMAE formulation prevents large errors in a few forecasts from dominating the 
error metric. 

The MASE, recommended by \citet{hyndman2006}, 
summarizes forecast performance in terms of the mean ratio
of errors from a proposed method in the numerator and errors from a na{\"\i}ve
method such as persistence forecast in the denominator. 
Error estimates in the denominator are estimated from mean errors of the na{\"\i}ve
method using the in-sample data, e.g. mean net migration rate errors generated from 
the 1950-2000 data using the na{\"\i}ve method in the denominator compared to mean 
forecast error in the 2000-2020 data. 
Let $f_{i,t}$ denote the net migration rate forecast for country $i$ and period starting 
in year $t$ when the true net migration rate is $r_{i,t}$.
The MASE of the forecast for $T_k$ forecast periods and $S_k$ in-sample periods starting in  
forecast period $t_0$ to $t_0+5$ and in-sample period $s_0$ to $s_0+5$ with a forecast 
horizon $k$ periods ahead of the last observation is defined as
\begin{equation}
    \text{MASE}_{k} = \frac{\frac{1}{200 \times T_k} \sum_{i=1}^{200} \sum_{\{t_0\}_k} \left| r_{i,t_0+5(k-1)} - f_{i,t_0+5(k-1)} \right|}{\frac{1}{200 \times S_k} \sum_{i=1}^{200} \sum_{\{s_0\}_k} \left| r_{i,s_0+5k} - r_{i,s_0} \right|}.
\end{equation}
This formula leads to the ratio of average $k$-period ahead forecast errors for all possible 
horizons $k$ in the in-sample and out-of-sample periods.
For the $k=1$ one-period-ahead forecasts, $\{t_0\}_1 = \{2000, 2005, 2010, 2015\}$, $T_1=4$, $\{s_0\}_1 = \{1950, 1955, \dots, 1990\}$, and $S_1=9$.  
For the $k=4$ four-period-ahead forecast, $\{t_0\}_4 = \{2000\}$, $T_4=1$, $\{s_0\}_4 = \{1950, 1955, \dots, 1975\}$, and $S_4=6$.

Forecast calibration was evaluated in terms of 95\% prediction interval coverage and 
average prediction half-interval width.
If a model is well-calibrated, then 95\% prediction intervals generated from the 
model should contain about 95\% of the true net migration rate observations. 
Prediction intervals that contain less than 95\% of the true values are too narrow, indicating that the model underestimates forecast variation. 
Prediction intervals that contain more than 95\% of the true values are too wide, 
indicating that the model overestimates forecast variation. 
Prediction half-interval width is calculated as half the difference between the 
upper prediction interval quantile and lower prediction interval quantile. 
Accurate, well-calibrated models with smaller half-interval widths are preferred
to models with the same accuracy and calibration estimates. 

Table \ref{tab:validation} summarizes out-of-sample predictive performance 
of the age-standardized, age-agnostic, and net rate persistence methods.
Errors are calculated as the difference between the observed value and the 
median of 2,000 posterior predictive distribution draws. 
Rate persistence uses the last observed net rate in each country as the forecast. 
The best score for each method is shown in bold font. 

\begin{table}[htb]
    \centering
    \caption{Mean predictive performance of different methods for net migration rate 
    (migrants per thousand period person years): mean absolute error (MAE), 
    log mean absolute error (LMAE) with $c=1$, mean absolute scaled error (MASE), 
    95\% prediction interval coverage, and mean prediction interval half-width (HW).}
    \begin{tabular}{llccccc}
        \toprule
                        & Method        & MAE           & LMAE              & MASE          & Cover         & HW\\
        \midrule
        5 Years         & Persistence   &4.02           &0.68               &1.42           &---            &--- \\
                        & Agnostic      &\textbf{3.44}  &\textbf{0.65}      &\textbf{1.25}  &\textbf{93}    &10.47 \\
                        & Standardized  &3.54           &\textbf{0.65}      &1.28           &\textbf{93}    &\textbf{10.39} \\
                        &               &               &                   &               &               & \\
        10 Years        & Persistence   &5.33           &0.88               &1.63           &---            &--- \\
                        & Agnostic      &\textbf{3.86}  &\textbf{0.76}      &\textbf{1.16}  &\textbf{91}    &11.63 \\
                        & Standardized  &3.93           &0.77               &1.17           &90             &\textbf{11.44} \\
                        &               &               &                   &               &               & \\
        15 Years        & Persistence   &5.16           &1.00               &1.44           &---            &--- \\
                        & Agnostic      &\textbf{3.49}  &0.83               &\textbf{1.10}  &\textbf{92}    &12.32\\
                        & Standardized  &3.51           &\textbf{0.82}      &1.11           &\textbf{92}    &\textbf{11.85}\\
                        &               &               &                   &               &               & \\
        20 Years        & Persistence   &4.77           &1.02               &1.27           &---            &--- \\
                        & Agnostic      &2.91           &0.82               &1.00           &\textbf{94}    &12.54\\
                        & Standardized  &\textbf{2.86}  &\textbf{0.79}      &\textbf{0.98}  &\textbf{94}    &\textbf{12.04}\\
        \bottomrule
    \end{tabular}
    \label{tab:validation}
\end{table}

The age-agnostic and net rate persistence models were fit using the bayesPop R 
package~\citep{sevcikova2016bayesPop}. 
The age-standardized model was fit using a custom implementation of the bayesPop R package. 
The age-standardized and the age-agnostic model accuracy were similar across all 
horizons, but age-standardized forecast accuracy overtakes age-agnostic forecast 
accuracy as the number of periods between the last observations and the forecast 
period increased for all metrics. 
The age-standardized model was better calibrated in terms of 95\% prediction interval 
coverage for all horizons except the one-period-ahead forecast, but both models had 
slight under-coverage compared to the 95\% nominal rate. 
The age-agnostic and age-standardized models outperformed the rate persistence model
forecast in all horizons.
Similarities between the agnostic and standardized model accuracy are expected over short 
forecast horizons since population age structure changes slowly.

\section{Results}\label{Results}

\begin{figure}[H]
    \centering
    \includegraphics[width=\textwidth]{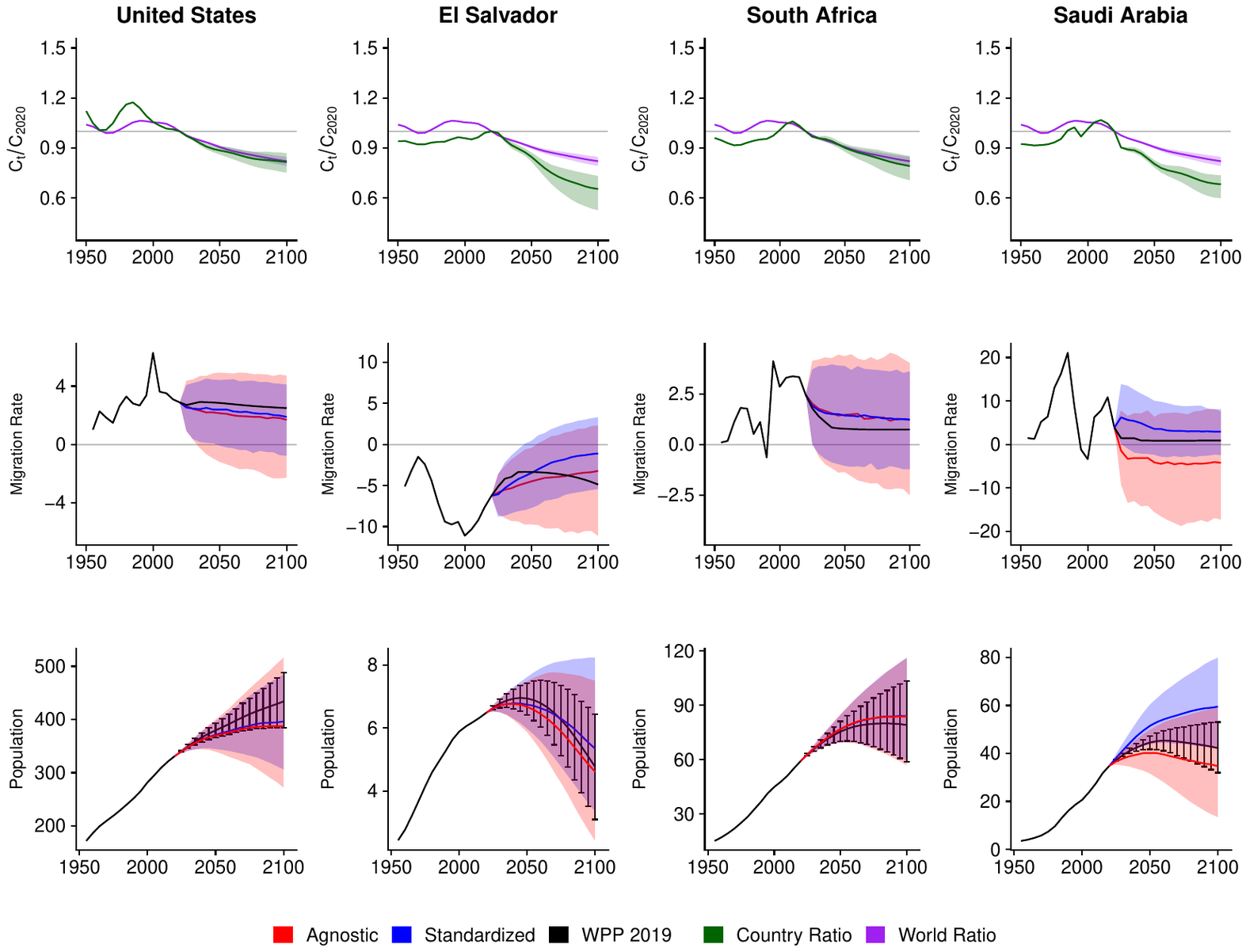} 
    \caption{2020 MASI ratios (top row), net migration rate as net annual migrants per thousand (middle row),
    and probabilistic forecast of population (in millions) age-standardized and age-agnostic (bottom row)
    for the United States, El Salvador, South Africa, and Saudi Arabia.
    Forecasts use probabilistic net migration (\protect\blue=age-standardized and \protect\red=age-agnostic), as well as probabilistic fertility, 
    and mortality. The MASI ratio plots show the values for each country (\protect\green) 
    and the world (\protect\purple). Solid lines in each plot indicate the observed and median forecast. 
    Shaded regions show the 80\% prediction interval. Forecasts start in the 2020-2025 period.}
    \label{fig:migPopforecast}
\end{figure}

Figure \ref{fig:migPopforecast} summarizes probabilistic population forecasts
in four countries using the age-agnostic migration model from \citet{azose2016}, 
the population forecasts using the age-standardized migration model, and 
the projections from the 2019 World Population Prospects (WPP 2019) \citep{unwpp2019}. 
Columns of the figure correspond to the same country. 
Rows of the figure show forecasts through 2100 for the MASI ratio, 
the net migration rate, and total population. 

The MASI ratios in the first row of Figure \ref{fig:migPopforecast} 
summarize the differences in the population age structure from the 2020 baseline
for each country (green) compared to the population age structure for the world (purple). 
Countries with a higher migration-age population than the 2020 population baseline rise above 1.
Countries aging faster than the world population fall below the global index.  
Forecast population age structures in the United States and South Africa generally follow the global 
aging trend. 

When the country-level MASI ratio is similar to the global one, then net migration 
forecasts from the age-standardized model and age-agnostic model should be more similar 
than in countries where the country-level and global MASI ratios diverge.
The population age structures of El Salvador and Saudi Arabia 
are forecast to age much faster than the global average. 
In these cases, the age adjustment shifts the migration forecast towards higher net inflows, as we see for El Salvador and Saudi Arabia.  

Median net migration forecasts from both probabilistic net migration models 
are similar in many countries, but long-term age-standardized net migration forecast intervals 
tend to be narrower than age-agnostic forecast intervals. 
This is explained by the population age normalization step used in the 
age-standardized method. 
Age-standardized net migration rate variation attributable to variation in historic 
population age structure is removed before fitting the \citet{azose2015} net 
migration model. 
Removing variation in past net migration estimates attributable to population 
age structure reduces the variance in the net migration model parameter estimates and hence 
the migration and population forecasts. 
Prediction intervals (80\%) for both probabilistic net migration models include the United Nations' 
WPP 2019 projections through 2100 for all countries shown in Figure \ref{fig:migPopforecast}. 

Median net migration forecasts from the age-agnostic and age-standardized methods are most 
similar in countries that have population age indices similar to the global population age index. 
The population age indices along with the median migration forecasts for the United States 
and South Africa in Figure \ref{fig:migPopforecast} demonstrate this trend. 
Age-standardized migration forecasts in countries where the population age 
structure diverges from the global norm often leads to much different median migration forecasts
compared to the age-agnostic migration model. 
For example, El Salvador's population is forecast to age much more rapidly from 2020 to 2100 
compared to the global age structure.  
As a result, the median age-standardized net migration forecast is shifted towards lower net 
out-migration compared to the age-agnostic forecast. 
Since the age adjustment leads to less negative net migration, the age-standardized population 
forecast is higher than the forecast using the age-agnostic model. 
Furthermore, the age-standardized 80\% prediction interval upper bound shows that 
there is less certainty that the population size will peak by 2100 compared to the age-agnostic 
model. 

Figure \ref{fig:migPopforecast} also shows that Saudi Arabia's population age structure is forecast to 
change about as fast as El Salvador's compared to the 2020 baseline population age 
structure. 
Saudi Arabia's net migration forecast is also shifted towards higher net in-migration, but 
this leads to a higher net positive migration forecast compared to the age-agnostic model. 
The age-standardized net positive migration forecast is better aligned to the historic 
net rate data in Saudi Arabia and the UN's WPP 2019 net migration forecast. 
Differences in Saudi Arabia's migration and population forecast using the age-standardized approach 
arise from both population age structure effects and adjustments to modeling unique features 
of migration in GCC countries. 
These adjustments lead to a substantially higher median population forecast and prediction interval. 
The UN WPP 2019 population forecast falls to the bottom of the age-standardized prediction interval 
compared to the upper middle of the interval using the age-agnostic model. 
See the supplemental index for detailed forecasts for all countries. 

\begin{figure}[H]
    \centering
    \includegraphics[width=\textwidth]{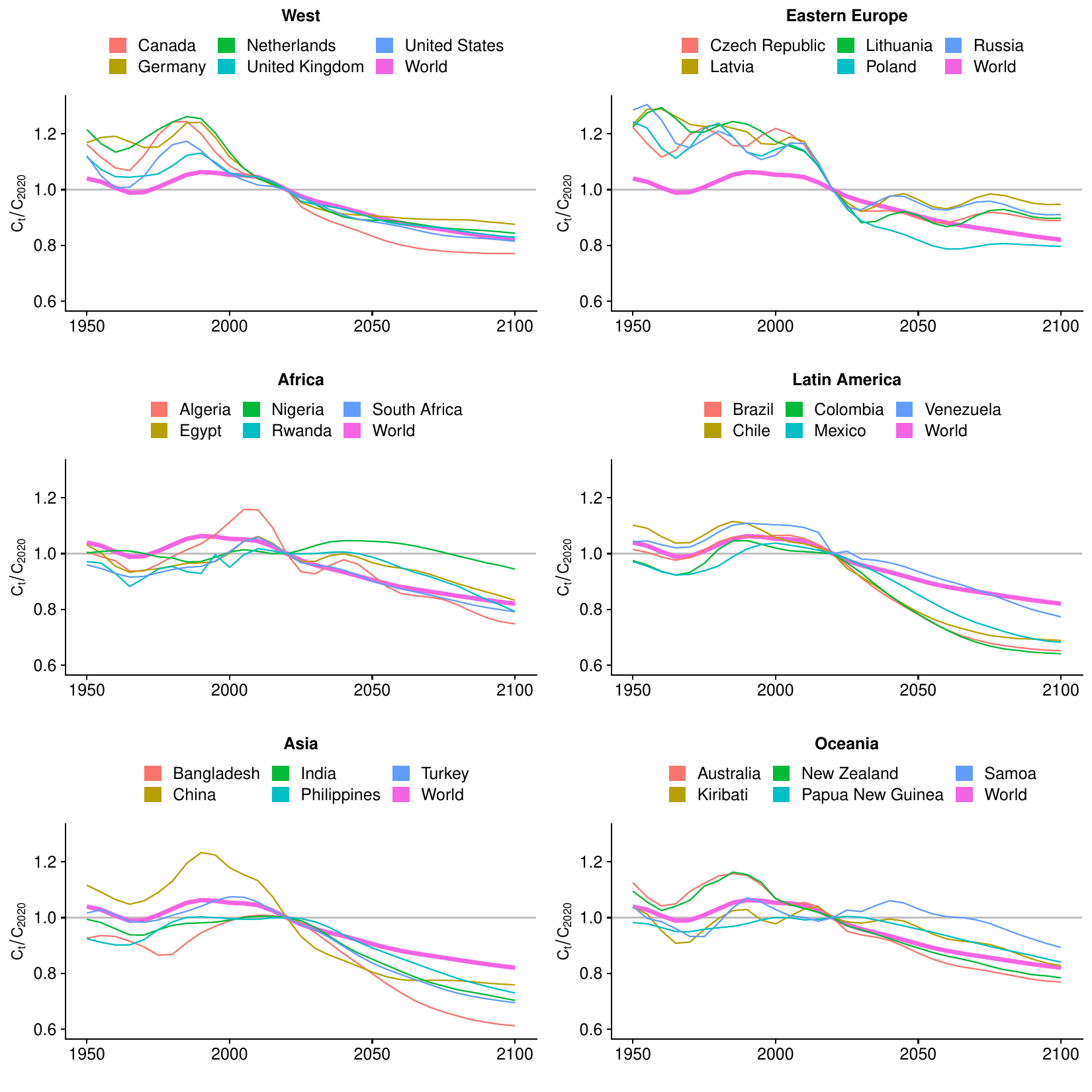} 
    \caption{Historic and median country-level forecasts of MASI ratios
    for 2020 baseline by region compared to the world.}
    \label{fig:ratioPlot}
\end{figure}

Figure \ref{fig:ratioPlot} summarizes past and median forecast MASI ratios for the globe and several countries by global region. 
The MASI ratio provides a concise one-number summary of historic 
and forecast population age structure dynamics. 
While mean country and global age also summarize the degree of aging, the 
MASI is more informative in terms of assessing 
changes to the inherent force of migration attributable to the relevant populations at risk. 
The MASI explicitly weights age groups by a Rogers \& Castro-like 
model migration age schedule. 
Taking the ratio of period MASI values puts the structural demographic force of migration 
potential within a country into the context of the reference population. 
The role of the MASI in removing the influence of population age 
structure is a more relevant measure of population age dynamics for migration 
than other measures that are less directly related to a population's 
structural migration potential, such as its average age.

Figure \ref{fig:ratioPlot} shows that the median population age structure forecasts in 
Western countries follows the global population aging trend.
This indicates that the underlying force of out-migration in Western countries is 
forecast to align with the global average compared to the population age structure 
in 2020 and the median net migration forecasts from the age-standardized and 
age-agnostic methods will largely agree with one another.  
The underlying force of migration from Latin America by contrast falls sharply in
coming decades compared to the 2020 baseline and the world overall, indicating 
that the force of out-migration will fall faster than the world on average and lead 
to higher net migration forecasts using the age-standardized method. 
The population age index forecast for some countries has yet to reach their peak, 
e.g. Rwanda in Africa and Samoa in Oceania. 
This implies that forecast out-migration from these countries is too 
small in the age-agnostic method. 

\begin{figure}[H]
    \centering
    \includegraphics[width=\textwidth]{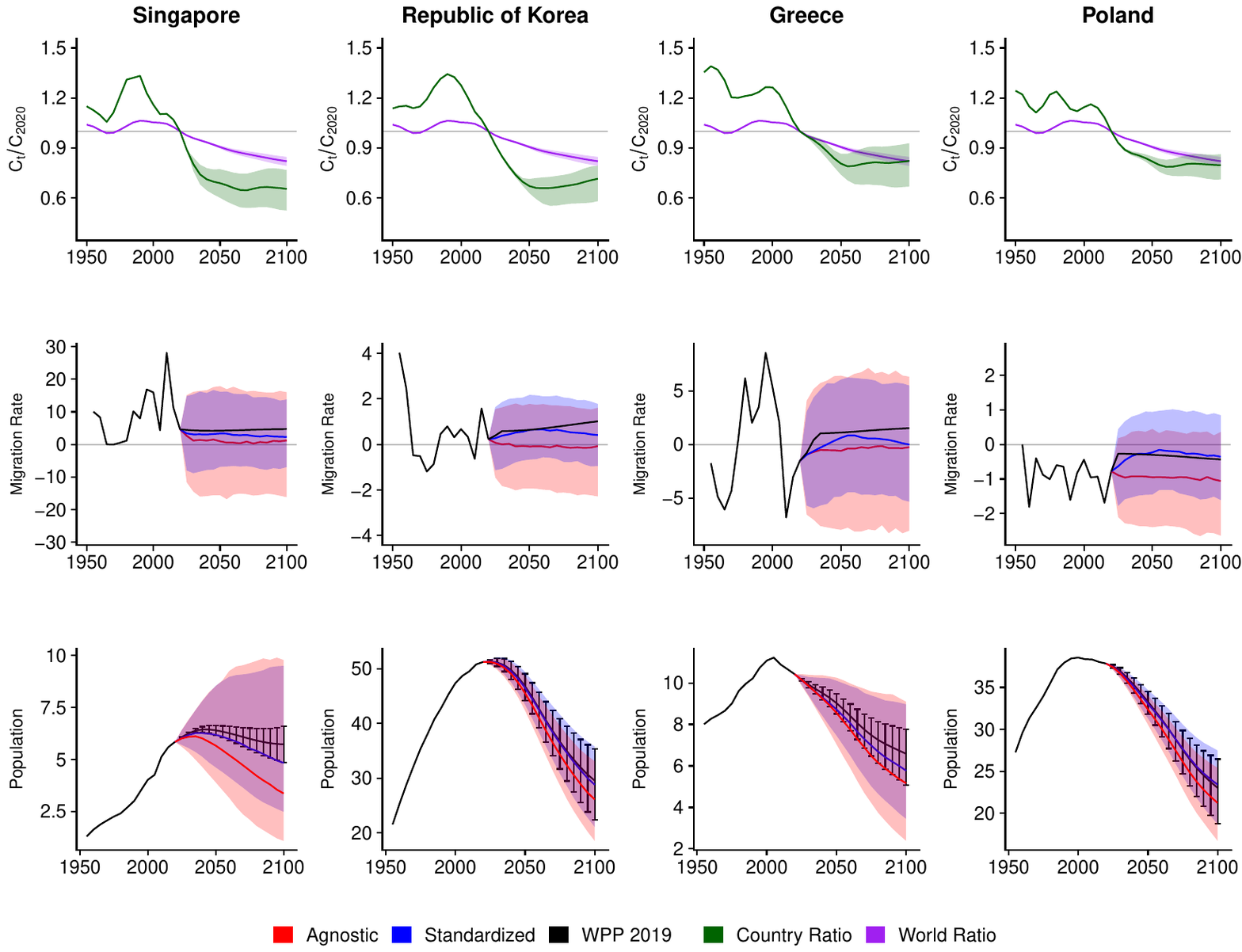} 
    \caption{2020 base-year MASI ratios (top row), age-standardized and age-agnostic net migration rate as net annual migrants per thousand (middle row),
    and probabilistic forecast of population (in millions) age-standardized and age-agnostic (bottom row)
    for Singapore, South Korea, Greece, Poland.
    Forecasts use probabilistic net migration (\protect\blue=age-standardized and \protect\red=age-agnostic), fertility, 
    and mortality. Age-index ratio plots show the age structure ratios for each country (\protect\green) 
    and world (\protect\purple). Solid lines in each plot indicate the observed and median forecast. 
    Shaded regions show the 80\% prediction interval. Forecasts start in the 2020-2025 period.}
    \label{fig:depop}
\end{figure}

Age-standardized migration forecasts led to less steep population declines 
in countries facing the greatest demographic challenges. 
Figure \ref{fig:depop} shows the age index, net migration, and population forecasts
among large countries that anticipate some of the most drastic population 
contractions by 2100. 
The age-standardized net migration forecast leads to less severe population forecast 
declines than the age-agnostic forecast as there are fewer people of prime migration 
age in the population. 
The age index shows that the force of out-migration falls in these countries as they 
age compared to the rest of the world, shifting the net migration forecast towards 
higher in-migration compared to the age-agnostic forecast. 
Median age-standardized net migration forecasts in these countries are more similar 
to the UN's WPP 2019 migration and population forecasts than the age-agnostic 
forecasts. 

\begin{figure}[H]
    \centering
    \includegraphics[width=\textwidth]{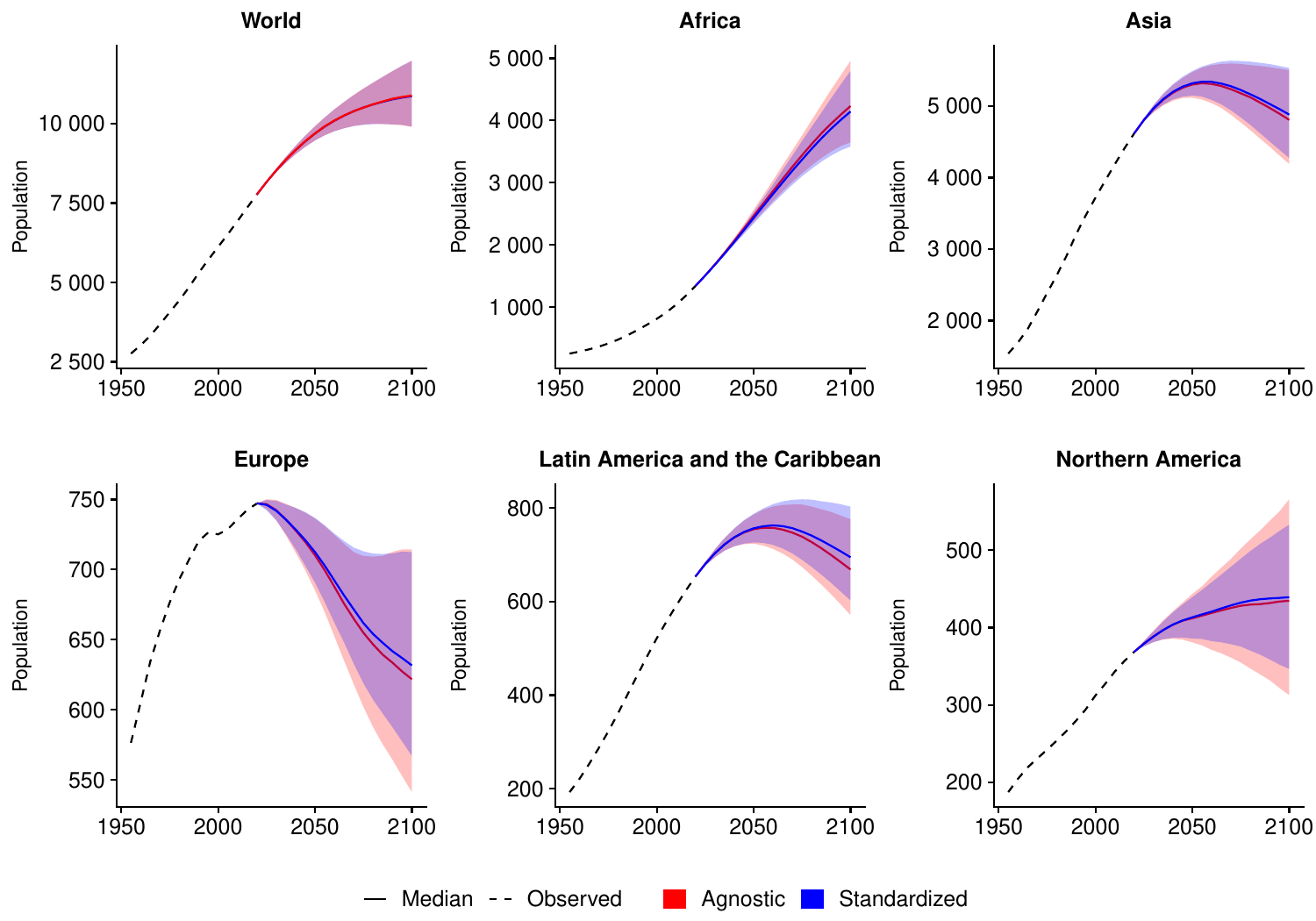} 
    \caption{Probabilistic forecast of population (millions of people) by region.
    Forecasts use probabilistic net migration (\protect\blue=age-standardized and \protect\red=age-agnostic), fertility, 
    and mortality. Solid lines indicate the median forecast. Dashed lines indicate the observed population.
    Shaded regions show the 80\% prediction interval. Forecasts start in the 2020-2025 period.}
    \label{fig:regionPop}
\end{figure}

Figure \ref{fig:regionPop} shows the world and regional forecasts using the age-agnostic and 
age-standardized migration forecast models.
World forecasts of population are indistinguishable, 
which makes sense as both methods use the same fertility and mortality models.
Global forecasts could have differed due to differences in fertility and mortality norms among migrants' destination countries, but 
these differences were not large enough to substantially alter the global population forecasts or overall uncertainty. 

Africa's population forecast by the end of the century is lower under the age-standardized model compared to the agnostic migration model. 
This makes sense because the force of migration associated with population age structure in many African countries is forecast to far exceed 
the global median for most of the remainder of the century.
As a result, long-term forecasts that ignore population age structure for many countries in Africa are slightly overstated compared to the age 
standardized migration model. 
End-of-century median population forecasts for all other regions are higher using the age-standardized migration model. 

The most pronounced differences are for Europe and North America. 
Age-standardized posterior prediction intervals for Europe and North America lie within the age-agnostic prediction intervals. 
The most extreme population declines in Europe and Latin America are less likely under the age-standardized model. 
Migration forecasts from the age-standardized model are lowered by aging populations in both regions. 
The most extreme population growth and decline trajectories in North America are less likely under the age-standardized model. 

Table \ref{tab:popmigdiff} details the differences in the population forecasts using the 
age-standardized and age-agnostic migration forecasting methods shown in Figure \ref{fig:regionPop}. 

\begin{table}[htb]
    \centering
    \caption{Median population and net migration forecasts at the end of the century by 
    United Nations Area and the world using age-standardized (S) and age-agnostic (A) 
    migration forecasting in millions of people along with the net difference (S-A) and 
    percent difference 100$\times \frac{S-A}{A}$ \%. }
    \begin{tabular}{llrrrr}
        \toprule
        UN Area                     & Measure                   & S     & A    & S - A  & 100$\times \frac{S-A}{A}$ \% \\
        \midrule
	Africa & population & 4142.6 & 4231.3 & -88.7 & -2.0 \\ 
  		& net migration & -10.5 & -6.3 & -4.3 & 68.0 \\ 
	& & & & & \\ 
  	Asia & population & 4880.3 & 4806.0 & 74.4 & 2.0 \\ 
  		& net migration & 0.1 & -1.2 & 1.3 & -109.0 \\ 
	& & & & & \\ 
  	Europe & population & 631.6 & 621.8 & 9.8 & 2.0 \\ 
  		& net migration & 3.4 & 2.8 & 0.7 & 24.0 \\ 
	& & & & & \\ 
  	Latin America & population & 694.4 & 668.3 & 26.2 & 4.0 \\ 
  		\& the Caribbean & net migration & 0.7 & -1.4 & 2.1 & -153.0 \\ 
	& & & & & \\ 
  	Northern America & population & 439.3 & 434.6 & 4.6 & 1.0 \\ 
  		&net migration & 4.4 & 3.9 & 0.5 & 12.0 \\ 
	& & & & & \\ 
  	Oceania & population & 65.1 & 62.6 & 2.5 & 4.0 \\ 
  	 	& net migration & 0.4 & 0.2 & 0.2 & 132.0 \\ 
	& & & & & \\ 
        \midrule
        World & population & 10863.3 & 10884.4 & -21.1 & 0.0 \\ 
        \bottomrule
    \end{tabular}
    \label{tab:popmigdiff}
\end{table}

\begin{figure}[H]
    \centering
    \includegraphics[width=\textwidth]{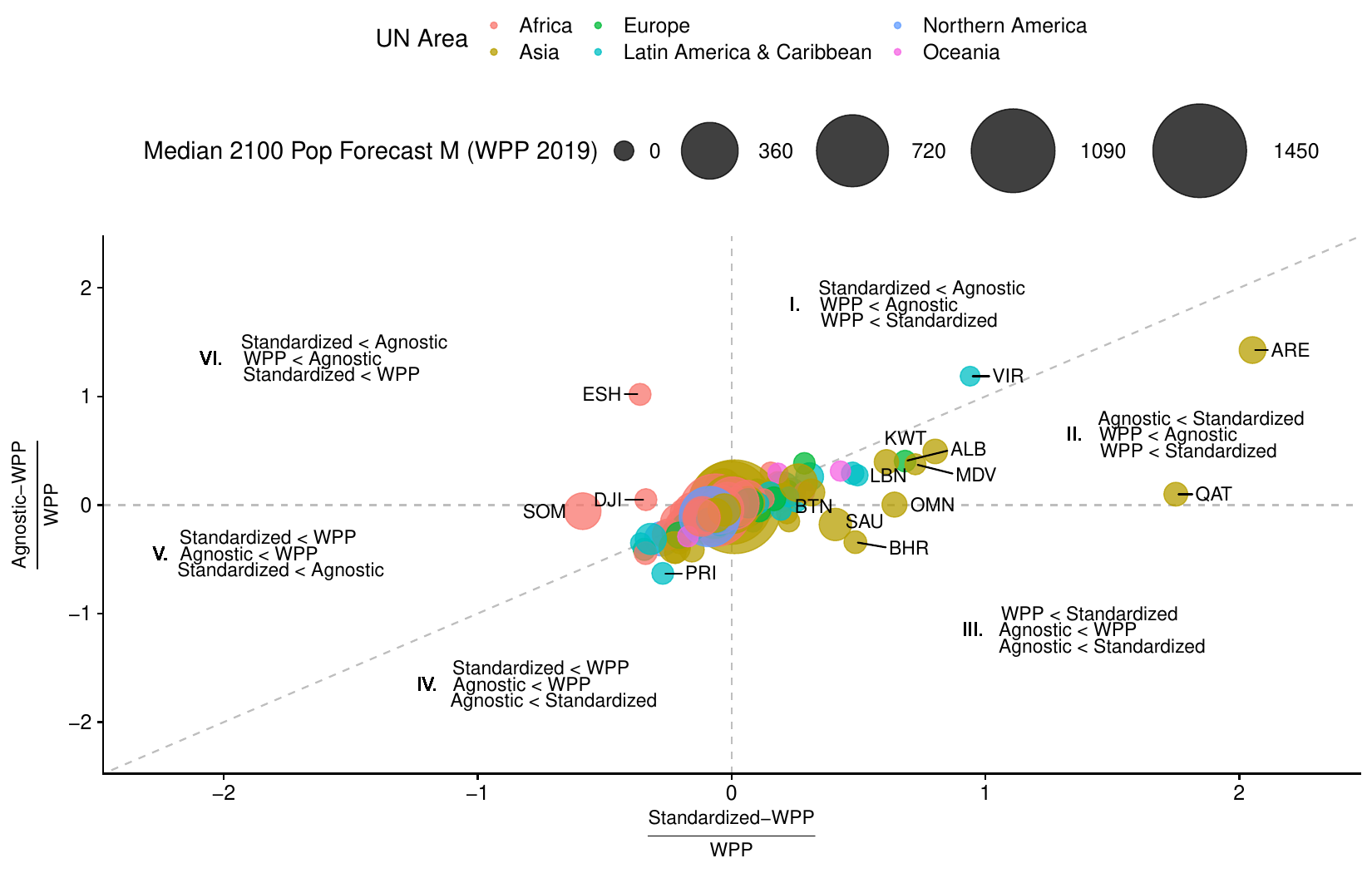} 
    \caption{Three-way comparison of median 2100 population forecasts using 
    age-standardized migration model, age-agnostic migration model, and median 
    WPP 2019 forecast. Point color indicates UN Area, point diameter indicates 
    WPP 2019 median population forecast size (millions), and regions I-VI show
    direction of forecast differences using the age-agnostic and age-standardized 
    migration models.}
    \label{fig:countrypopdiff}
\end{figure}

Figure \ref{fig:countrypopdiff} shows the population forecast differences at 
the end of this century for each country for three different forecasts.
Each point denotes the departure of the median population forecast by 2100 
using the age-agnostic or age-standardized migration model from the median 
WPP 2019 forecast. 
These ratios make it possible to evaluate differences in all three forecasts 
together. 
Points falling near the origin indicate approximate agreement between the 
age-agnostic, age-standardized, and WPP 2019 median population forecasts. 
Points on along $y=0$ indicate agreement between median age-agnostic and 
WPP 2019 forecasts that differ from the age-standardized forecast. 
Points on $x=0$ indicate agreement between the age-standardized forecast 
and the WPP 2019 forecast that differ from the age-agnostic forecast. 
Points on $y=x$ far from the origin denote agreement between the age 
agnostic and age-standardized forecasts that depart from the WPP 2019
forecast. 
All other combinations are summarized by the plot regions bounded by 
the dashed lines and labeled I-VI on the plot. 

Figure \ref{fig:countrypopdiff} shows that all three median population forecasts 
differ for Eastern Sahara (ESH). 
Both the age-agnostic and age-standardized methods lead to higher population forecasts 
for the United Arab Emirates (ARE) than WPP 2019, but the median age-standardized 
forecast is largest. 
The median age-standardized population forecast was highest among the three for 
all GCC countries. 
The age-standardized method generated fewer negative net migration trajectories 
over the forecast period; however, this outcome is consistent with the last three 
decades of sustained positive net migration that has been dominated by foreign workers 
responding to labor opportunities in rapidly expanding economies. 
Net migration rate forecasts using the age-standardized approach continue the 
net positive trend in GCC countries with expanding populations. 
While the age-standardized migration forecasts for GCC countries is higher than 
the age-agnostic approach, the population age structure in most GCC countries 
appears much more plausible and closer to the WPP 2019 forecast than the age-agnostic method. 

The Supplementary Information includes forecasts for the 200 largest countries, associated quantiles, 
and additional information about the software used to generate these results.

\section{Discussion}\label{Discussion}

The association between migration and age is long established. 
However, we are unaware of any existing migration forecasting method that explicitly 
accounts for the influence of overall population age structure, apart from
that of \citet{raftery2023}, which relied on an unrealistic assumption.
The potential support ratio (PSR), defined as the proportion of the working age population 
to the total population, has been used as a proxy for population age structure in 
other methods \citep{kim2010}. 
However this approach does not distinguish a population age distribution 
concentrated at ages 20-39 from one more heavily concentrated at 
ages 40-65, although the implications of these for migration are very different.

Our approach addresses multiple criticisms of using net migration as a unit 
of analysis. 
Our model-based approach to decompose the net flow rate into inflow and outflow
rates alleviates the migration age schedule issues outlined in \citet{rogers1990}. 
He also pointed out that relatively small net migration counts over a 
period can hide comparatively huge inflows and outflows of roughly equal magnitude. 
While net migration is the unit of analysis here, we use the historic data within 
a country to decompose net flow rate estimates into inflow and outflow rates. 
Dividing inflow and outflow counts by the age-standardized population size makes 
it possible to put flow rates from a diverse population of countries around the globe 
on the same scale for modeling and forecasting. 

While the methods used in age-standardized net migration estimation and forecasting partly
address long-standing criticisms of net migration rate analysis, a bilateral migration flow 
model among all countries as in \citet{welch2022} eliminates theoretical challenges induced 
by the use of net migration rate as the unit of analysis. 
However, the computational complexity of migration flow forecasting and limited 
historic bilateral flow data reduce the viability of this alternative for longer-term forecasting. 
International bilateral migration flow forecasting among 200 countries involves 
estimating and forecasting 39,800 bilateral flows as opposed to 200 net flow rates. 
Furthermore, there are only six quinquennial periods of bilateral migration flow 
estimates available at present \citep{abel2019} compared to thirteen periods 
for net migration \citep{unwpp2019}. 
Methods described in this article offer a more computationally 
tractable alternative to bilateral migration flow model 
fit to only half the historical net migration data. 

Probabilistic forecasts offer a principled means to integrate uncertainty, but 
historic shocks to long-term trends arise without warning. 
Multiple such shocks occurred in the last ten years alone.
The 2020 COVID-19 pandemic, Syria's decade old civil war, Russia's invasion in 
Ukraine, and the economic meltdown in Venezuela all upset long-term demographic 
norms---especially migration. 
New norms could take hold as a result of these and other crises, but the 
full effect of these historic events should be integrated into updated forecasts 
once the data become available. 
The age-standardized probabilistic migration forecasting method presented here 
includes well-defined mechanisms to account for changes to variation around long-term trends. 
When the full extent of recent crises emerges in the data, the forecasts should be 
revised with these data. 

Using age-standardized net migration aims to eliminate the influence of population age 
structure on historic and forecast net migration rates, improving on existing 
probabilistic net migration forecasting methods. However, key drawbacks of 
the use of net migration are not completely addressed by our method. 
Shifts in net migration forecast due to changes in country population age structure 
are directly reflected in the forecast for the country of origin; however, 
using the population age structure of the globe to normalize inflow 
rates could be improved to more directly reflect the changes in population age structures
of sending countries. 
For example, Mexico and Central American countries are currently the largest suppliers of immigrants to the United States. 
Calculating a country-specific inflow age ratio index to reflect that fact rather than 
a simple average of the global population age distribution would more directly link changes
in the population age structure of top sending countries on the age-standardized net migration rate.
We chose to use the global average in our models because it is far simpler 
analytically and to account for the possibility that 
migration corridors that exist now could change in the future.

Our methods build on methodological advances made in the last decade 
\citep{azose2015}, address several known challenges with net migration as the unit of analysis 
\citep{rogers1990}, and offer a more computationally tractable alternative to migration 
flow forecasting \citep{welch2022}. 
Our model-based approach to decomposing net migration rates into inflow and 
outflow rates offers an efficient alternative to more computationally challenging 
analyses involving direct modeling of bilateral migration flows. 
We propose a new method to account for unique differences in forecasting 
net flows in Gulf Cooperation Council countries that more accurately reflects 
migration dynamics in those countries in the future. 
Finally, we found the MASI ratio to be a concise one-number summary 
of the population age structure that measures the pace of 
aging expected to take place in the coming decades in the context of migration.
Reducing the influence of population age structure on historic and 
forecast migration rates leads to more accurate long-term forecasts and matches 
or narrows the migration and population prediction intervals compared to a comparable 
age-agnostic method. 

\nolinenumbers

\bibliography{bibliography} 

\newpage 

\appendix

\section{Proof of Theorem 1}
\begin{proof}
Starting from the definition of $\mbox{OMR}_{i,t}$, 
\begin{align}
\mbox{OMR}_{i,t} &= \frac{O_{i,t}}{\tilde{P}_{i,t,+,+}} & \left( \mbox{definition of OMR} \right) \nonumber \\
    &= \sum_a \frac{O_{i,t,a}}{\tilde{P}_{i,t,a,+}} \times \frac{\tilde{P}_{i,t,a,+}}{\tilde{P}_{i,t,+,+}} & \left( \mbox{OMR disaggregated by age} \right) \nonumber \\
    &= G_{i,t} \sum_a \frac{\mbox{OMR}_{i,t,a}}{G_{i,t}} \times \pi_{i,t,a} & \left( \mbox{Multiplication by } \frac{G_{i,t}}{G_{i,t}} \right). \nonumber \\
\end{align}
The same derivation applies to $\mbox{OMR}_{i,t}^*$, implying that
\begin{equation}
\mbox{OMR}_{i,t}^* = G_{i,t} \sum_a \frac{\mbox{OMR}_{i,t,a}}{G_{i,t}} \times \pi_{i,t,a}^*. \label{line-OMR*}
\end{equation}
Rearranging $\mbox{OMR}_{i,t}$ in terms of $G_{i,t}$ and substituting the equation into (\ref{line-OMR*}) yields the desired result:
\begin{equation}
 \mbox{OMR}_{i,t}^* = \mbox{OMR}_{i,t} \frac{\sum_a \mbox{OMR}_{i,t,a} \pi_{i,t,a}^*}{\sum_a \mbox{OMR}_{i,t,a} \pi_{i,t,a}}. \label{line-result}
\end{equation}
Multiplying (\ref{line-result}) by $\frac{G_{i,t}}{G_{i,t}}$, assuming the same $R_a = \mbox{OMR}_{i,t,a}/G_{i,t}$ in both populations, and applying the definition of $R_a$, we obtain
\begin{equation}
\mbox{OMR}_{i,t}^* = \mbox{OMR}_{i,t} \frac{\sum_a R_a \pi_{i,t,a}^*}{\sum_a R_a \pi_{i,t,a}} = \mbox{OMR}_{i,t} \frac{C^*_{i,t}}{C_{i,t}},
\end{equation}
as desired. This concludes the proof.
\end{proof}

\end{document}